\documentclass[twocolumn,groupedaddress,amsmath,aps,pra]{revtex4-1}

\usepackage{graphicx}

\begin{document}
\title{Laser-induced atomic adsorption: a mechanism for nanofilm formation}

\author{Weliton S. Martins, Thierry Passerat de Silans, Marcos Ori\'{a} and Martine Chevrollier}
\email{martine@otica.ufpb.br}
\affiliation{Laborat\'{o}rio de Espectroscopia \'{O}tica, DF-CCEN, Cx. Postal 5086 - Universidade Federal da Para\'{i}ba, 58051 - 900 Jo\~{a}o Pessoa - PB, BRAZIL}

\begin{abstract}
We demonstrate and interpret a technique of laser-induced formation of thin metallic films using alkali atoms on the window of a dense-vapour cell. We show that this intriguing photo-stimulated process originates from the adsorption of Cs atoms via the neutralisation of Cs$^+$ ions by substrate electrons. The Cs$^+$ ions are produced via two-photon absorption by excited Cs atoms very close to the surface, which enables the transfer of the laser spatial intensity profile to the film thickness. An initial decrease of the surface work function is required to guarantee Cs$^+$ neutralisation and results in a threshold in the vapour density. This understanding of the film growth mechanism may facilitate the development of new techniques of laser-controlled lithography, starting from thermal vapours.
\end{abstract}

\maketitle

Atomic vapours in optical cells are convenient and ubiquitous systems in many studies of matter-radiation interaction. Interactions of the atomic samples with the cell walls are often unavoidable and may interfere with the processes studied, because of the production of stray electric fields \cite{Obrecht2007A}, spin depolarization \cite{Happer1972}, surface conductivity \cite{Bouchiat1999} and optical transparency impairment due to the reactivity of hot alkali atoms with the glass. On the other hand, these atom-surface interactions can be explored for applications in fundamental as well as in applied physics. For example, alkali adsorbates represent a controllable source of atoms via light-induced desorption \cite{Hatakeyama2006}. Selective, laser-controlled quantum adsorption of cold atoms for realisation of 2D atomic waveguides remains a theoretical possibility \cite{Lima2000,deSilans2006,leKien2007}, to be experimentally demonstrated. The perspective of controlling adsorption using light opens the way to grow nanostructures directly on a dielectric surface, which is particularly attractive for such applications as building circuitry with the aim of manipulating cold and ultracold atomic samples close to surfaces \cite{Gunther2005}. We demonstrate here the laser-induced adsorption of caesium atoms from a dense hot vapour, leading to the formation of a thin caesium film on a dielectric window when illuminated with a laser resonant with the Cs D$_2$ transition. The occurrence of this photo-stimulated atom-surface process has essentially gone unnoticed or ignored, in spite of its potential relevancy in studies involving alkali vapour cells \cite{Jha2011} or as a new lithographic technique \cite{Afanasev2008}, for which the fundamental mechanisms have been elusive until now \cite{Afanasiev2007}.

\begin{figure}
\includegraphics{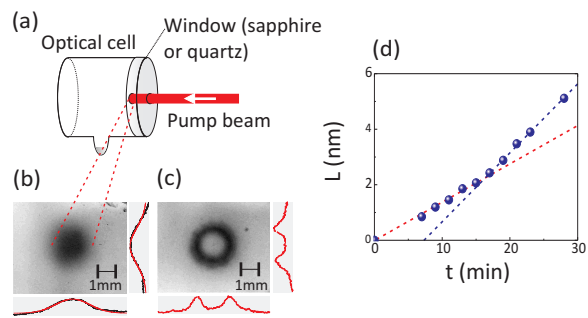}
\caption{(color online). (a) Experimental set-up. Laser power $\approx$ 10 mW, waist $\approx$ 1.5 mm, frequency detuning $|\delta| <$ 1 GHz. (b,c) Image of a thick caesium film (greyscale) with thickness (red profiles) that follows the laser beam transverse intensity profile. For illustrative purpose, the beam have been given (b) a Gaussian and (c) a ring shape. (d) Metallic film thickness increasing as a function of time.}
\label{fig:setup}
\end{figure}

The basic experimental configuration shown in fig.~\ref{fig:setup}(a) consists in a low-power (a few mW/mm$^2$, beam radius $\approx$ 1 mm) cw laser beam sent through an optical cell containing a hot vapour of caesium atoms. The laser frequency is tuned close to the frequency of the alkali D$_2$ transition ($\lambda$ = 852 nm). Surprisingly, we observe the formation of a metallic film on the illuminated spot at the interface between the window of the cell and the vapour \cite{Afanasiev2007}. The thickness profile of the film follows the transverse intensity of the pump laser beam (figs.~\ref{fig:setup}(b) and \ref{fig:setup}(c)), which suggests potential applications as a lithographic technique \cite{comentario1}. Light-induced formation of CsH particles in metal vapours and their subsequent deposit as a dielectric film has been reported in the literature \cite{Tam1975} but the phenomenon observed and described in the present work is not related to this so-called 'laser snow', and  the films we study are clearly metallic \cite{Afanasiev2007}. We can infer the average thickness, $L$, of a film with Gaussian profile from the transmission of a non-resonant (NR), low-power He-Ne laser probe ($\lambda_{NR}$ = 632.8 nm, $I_{NR}\approx$ 0.6 mW.cm$^{-2}$) through the optical cell. We assume this transmission follows the Beer-Lambert law, $I(L)/I_{NR}= \textrm{exp}(-4 \pi/\lambda_{NR})n'' L$, where $n''$ is the imaginary part of the index of refraction of the cesium film. We use the bulk extinction coefficient of cesium, $n''$= 1.28 at 633 nm \cite{Palik1985}. The pump beam is blocked during measurements of the probe-beam transmission through the cell. The ratio of probing to growing time is typically $1/300$. We measure a film growth rate on the order of $10^{-3}$ nm.s$^{-1}$.

\begin{figure}
\includegraphics{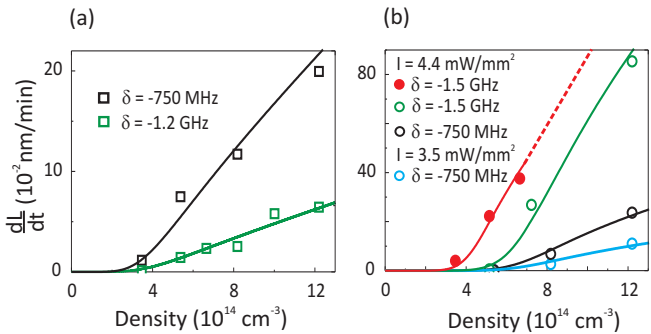}
\caption{(color online). Growth rate of the light-induced film in the initial few-layers regime, as a function of the atomic vapour density: (a) in a sapphire window, for $I$ = 3.2 mW.mm$^{-2}$ and $T_w$ = 215 $^\circ$C; (b) in a quartz window. Open circles: $T_w$ = 215 $^\circ$C; solid circles: $T_w$ = 190 $^\circ$C. Solid lines: fits obtained from eqs.~(\ref{equ:rate})-(\ref{equ:p}).}
\label{fig:density}
\end{figure}

\begin{figure}
\includegraphics{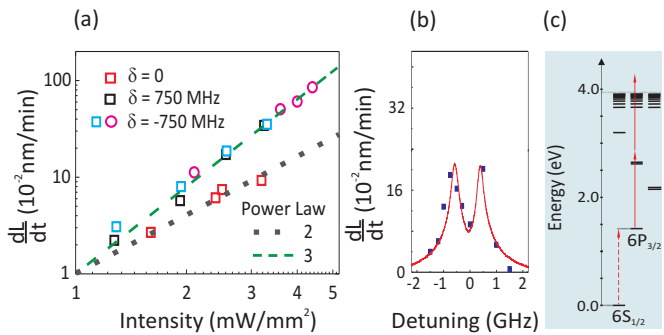}
\caption{(color online). (a) and (b) Growth rate of the light-induced film in the initial few-layers regime. (a) as a function of the laser intensity for $n$ = 12 $\times 10^{14}$ cm$^{-3}$ and $T_w$ = 215 $^\circ$C. Circles: quartz window; squares: sapphire window. All of the rates follow a power law $dL/dt = A I^{\mu}$, where $\mu$ = 2,3 (dotted and dashed lines, respectively). The curves have been normalised to $A$ = 1. (b) as a function of laser detuning from the Cs D$_2$ line center for a sapphire window at $T_w$ = 215 $^\circ$C, $n$ = 12 $\times 10^{14}$ cm$^{-3}$ and $I$ = 3.2 mW.mm$^{-2}$. The solid line is the spectrum of the back-scattered fluorescence at 852 nm (see text and \cite{Zajonc1981}). The relative uncertainties are $\approx 20$\% on the intensity, $\approx 15$\% on the growth rate and $\approx 5$\% on the laser detuning. (c) Relevant Cs energy levels and the ionization scheme of the excited atoms.}
\label{fig:intensity}
\end{figure}

The metallic film induced by laser is characterized by two distinct growth rates (see fig.~\ref{fig:setup}(d)). The typical growth rate change observed at $t \approx$ 15 min occurs when the \textit{average} film thickness is on the order of a few atomic layers. We study the film formation in the \textit{initial} regime:
Figures \ref{fig:density}(a) and \ref{fig:density}(b) show the behaviour of the film growth rate as a function of the atomic density in the vapour for two different dielectric substrates (sapphire and quartz, respectively) and for different sets of window temperature, laser frequency detuning and power.  We finely control the temperature of the cell windows and of the reservoir. The latter determines the vapour density and is set at a few tens of $^\circ$C colder than the cell body. The temperature of the reservoir ranges between 150 $^\circ$C and 190 $^\circ$C, corresponding to atomic densities between 2 $\times 10^{14}$ and 1.2   $\times 10^{15}$  cm$^{-3}$. The measurements have been made at two different temperatures of the window, 190 $^\circ$C and 215 $^\circ$C. The temperatures are stabilised with a precision of 1 $^\circ$C, leading to a relative uncertainty of $\approx$ 4 \% in the atomic density. The uncertainty on the growth rate is $\approx 15$\%. 

The growth rate is linear with atomic density, suggesting a one-atom process. However, there is a peculiar density threshold for film formation. All other parameters being equal, this threshold is larger for quartz (fig.~\ref{fig:density}(b)) than for sapphire (fig.~\ref{fig:density}(a)) surfaces, increases with the temperature of the window (fig.~\ref{fig:density}(b)) and does not depend on pump power or frequency (figs.~\ref{fig:density}(a) and \ref{fig:density}(b)), which only affect the efficiency of the film growth. 

	The growth rate increases cubically with the pump laser intensity (fig.~\ref{fig:intensity}(a)) for a frequency detuned from resonance so as not to saturate the atomic transition (detunings on the order of the transition Doppler linewidth). Thus, a three-photon process is the basis of the film growth on the surface. Notice that three 852-nm photons have a total energy of approximately 4.37 eV, sufficient to ionize a caesium atom (ionization energy $\approx$ 3.89 eV). This cubic dependence has been verified for various detunings on both sides of the atomic one-photon resonance, in the non-saturating regime (fig.~\ref{fig:intensity}(a)). Complementarily, using two lasers of similar frequencies, intensities $I_1$ and $I_2$ and with no fixed phase relationship, we studied the role of laser coherence effects on the film formation. Because the lasers frequency is close to resonance with the Cs D$_2$ transition, it is expected that, in a first step, the excited level 6P$_{3/2}$ will be significantly populated by any of the lasers, with a probability proportional to the total intensity. Indeed, series of experiments on the film formation using different combinations of the intensities $I_1$ and $I_2$ pointed to the interpretation that the growth rates are only compatible with the dependence expected for coherent two-photon absorption by 6P$_{3/2}$ excited atoms, in both the resonant and the non-resonant regimes (see figs.~\ref{fig:intensity}(a) and \ref{fig:intensity}(c)). This interpretation is further supported by the observation that the growth rate (see fig.~\ref{fig:intensity}(b)) follows the 6P$_{3/2}$ fluorescence spectrum, which exhibits a typical minimum at resonance, due to excitation quenching at the surface \cite{Zajonc1981}. Moreover, for small detunings, the laser saturates the D$_2$ transition and the two-photon absorption from the saturated 6P$_{3/2}$ excited state results in a growth rate increasing quadratically with the beam intensity (fig.~\ref{fig:intensity}(a)).

Therefore, our measurements show that, in the early stages of film formation, there is a linear growth with vapour density and a cubic dependence on the laser intensity. These are our main results, which demonstrate that, among many multiple-atom and multiphoton processes near the surface, the film genesis is due to single-atom ionization with three photons. This is a low-probability mechanism, which explains the relatively slow process of alkali film formation on windows of vapour cells in the presence of low-power resonant lasers. The atomic-density threshold further limits the conditions in which such a laser-induced atom-surface process can be observed. Therefore, this process is not ordinarily observed in low-density experiments.

To explain these observations, the processes that occur at the interface are described using simplified rate equations, in terms of adsorption and desorption rates. The interaction between atoms and surfaces originates either from the dipolar van der Waals interaction (physisorption, see for instance \cite{deBoer1953,Hoinkes1980}) or from charge exchange with the substrate (chemisorption, see for instance \cite{Norskov1990}). Pure physisorption and chemisorption constitute limiting cases and varying degrees of hybridization can actually be present. Adsorption of neutral atoms from the thermal vapour acknowledgedly falls into the physisorption category, while the light-induced film, intermediated by ions, involves stronger charge exchange processes and falls into the chemisorption one. These two processes, as expected, evolve at very different rates: at the typical temperatures we worked, we measured the desorption rate of the thermal atoms to be on the order of 10$^3$ s$^{-1}$, while the desorption rate of the film (neutralised ions) is on the order of 10$^{-3}$ s$^{-1}$ \cite{comentario2}. 
	
	The evidence for very different time scales for the two processes in our system allows us to separately consider the contributions from the neutral atoms in the vapour and from the neutralised ions. Therefore, for the very initial regime of film formation, the two equations can take the same general form:
	
\begin{equation}
\frac{dN_{j}}{dt} = p_{j} \; \sqrt{\frac{k_B T_w}{2 \pi m}}\;n_{j}  -\frac{1}{\tau_{j}} \; e^{-E_{j}/k_B T_w}\;N_{j},
\label{equ:rate}
\end{equation}	

where the subscript index $j=A,I$ stands for the thermal neutral atoms ($A$) and the neutralised ions ($I$). $N_{j}$ is the surface density of the $j$-kind adsorbed atoms (adatoms) in the area illuminated by the laser, $p_{j}$ is a sticking coefficient, $n_A$ ($n_I$) is the atomic (ionic) density in the vapour, $k_B$ is the Boltzmann constant, $m$ is the atom mass and $T_w$ is the window temperature. The first and second terms in eq.~(\ref{equ:rate}) are the adsorption and an Arrhenius-like \cite{Laidler1996} desorption rate per unit area, respectively. The residence time of the $j$-adatoms in the surface potential well is characterized by the adsorption energy $E_{j}$ and by the time constant $\tau_{j}$, associated to physisorption ($j=A$) and chemisorption ($j=I$) bound states. The adatom surface density in eq.~(\ref{equ:rate}) is proportional to the measured average film thickness $L$, through a coefficient $N_{ml}/(2r)$, where $r$ is the cesium atom radius ($r$ = 0.26 nm) and $N_{ml}$ the  cesium surface density for a monolayer. 

Before the light is turned on, the physisorbed atoms are in thermal equilibrium with the surface, forming a submonolayer film, i.e., the fluxes of thermal particles going to and from the surface are taken as being the same. Although they do not directly participate in the film growth, their equilibrium density on the surface determines the effective substrate for the light-induced ions incident from the vapour. Let us therefore first analyse the thermal equation rate (eq.~(\ref{equ:rate}) for $N_A(t)$). The equilibrium density of adatoms $N_{A0}$ on the surface results from the balance between the desorbed and adsorbed neutral atomic fluxes on the surface and is obtained by solving the thermal rate equation (\ref{equ:rate}) for $dN_A/dt=0$:
\begin{equation}
 N_{A0}(T_w,n_A)=p_{A} \; \sqrt{\frac{k_B T_w}{2 \pi m}} \tau_A\; e^{E_A/k_B T_w}\;n_A.
 \label{equ:Nab}
\end{equation}
It is responsible for the density threshold observed in figs.~\ref{fig:density}(a) and \ref{fig:density}(b). Alkali adatoms are known to significantly lower the work function of metal substrates and are used to increase surface ion or electron emission \cite{Yu1984}. Studies of the interaction of alkali atoms with insulators are less numerous but confirm that alkali adatoms also lower the work functions of dielectric substrates \cite{Brause1997}. 

\begin{figure}
\includegraphics{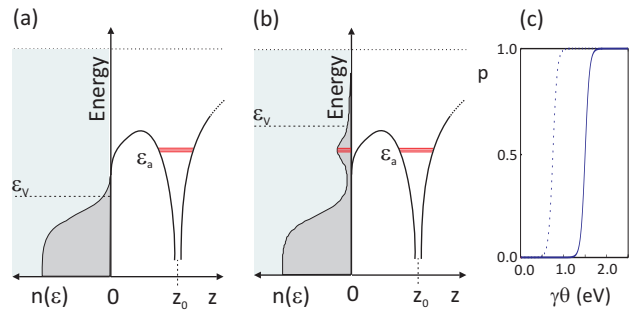}
\caption{\label{fig:estates} (color online).(a,b) Left-hand side: Evolution with Cs coverage of the electronic surface density of states of a dielectric substrate. $\epsilon_v$ is the energy of the valence-band edge. Right-hand side: electronic potential for a Cs atom located at distance $z_0$ from the plane insulator surface. $\epsilon_a$ is the atomic valence state energy. As Cs coverage increases, from 0 in (a) to $\theta > 0$ in (b), electronic states emerge in the substrate at the energy of the atomic valence state, which raises the valence-band edge and decreases the substrate work function from $\phi_0$ to $\phi(\theta)$. (c) Evolution with caesium coverage of the probability of ion neutralisation $p_I$ at the surface, for $\phi_0-\epsilon_a$= 0.75 eV (dotted line) and $\phi_0-\epsilon_a$= 1.5 eV (solid line), considering a linear decrease of the work function with coverage, $\phi=\phi_0-\gamma \theta$.}
\end{figure}

The interaction between atoms or ions and a partially caesiated solid surface depends on the relative electronegativities of the surface and the impinging species \cite{Kamins1968a}. The relative electronegativity, in turn, depends on the pre-existent adatom coverage $\theta$ (see fig.~\ref{fig:estates}). Coverage is defined as the ratio between the density of adatoms and the density $N_{ml}$ for a monolayer. At low Cs coverage (fig.~\ref{fig:estates}(a)), the surface is more electronegative than the Cs atoms, and thus, the impinging Cs atoms transfer their valence electron to the substrate \cite{Brause1997}. The positive core of adatoms then gives rise to an \lq\lq electrical double layer\rq\rq on the surface \cite{Gurney1935}. A higher coverage of Cs adatoms lowers the work function of the solid substrate \cite{Brison2007} and thus diminishes the relative electronegativity between the adatoms and the surface (fig.~\ref{fig:estates}(b)). 

As a result of the thermal surface caesiation, two interaction regimes may therefore occur between the caesiated surface and impinging Cs$^+$ ions. If the Cs coverage is low, the ions are repelled from the surface by the adsorbed atoms and their outward positive pole, thus preventing the film from forming. If the Cs coverage is sufficiently high to lower the substrate work function to less than the atom ionization energy, an electron can be transferred from the substrate to the ion, thus neutralising it \cite{Winter2002}. The vapour density threshold, which varies with the window temperature, therefore corresponds to a thermal coverage threshold. Of course, this threshold also depends on the window material, as shown in figs.~\ref{fig:density}(a) and \ref{fig:density}(b). For a given material, the variation of the threshold coverage with temperature is given by eq. (\ref{equ:Nab}): if the temperature $T_w$ increases, the atomic density $n_A$ must increase to compensate the decreasing term $\sqrt{T_w}\;\textrm{exp}(E_A/k_B T_w))$ and maintain the density of adatoms constant. This is what we observe in fig. \ref{fig:density}(b), where the threshold at $T_w$ = 215 $^\circ$C is obtained for an atomic density larger than the one necessary to reach the threshold at $T_w$ = 190 $^\circ$C.

We now analyse in more details the radiative equation (eq.~(\ref{equ:rate}) for $N_I(t)$). Unlike the thermal neutral atoms, the ion flux toward the surface is very different from the flux of the `chemisorbed' atoms escaping from the surface: as stressed before \cite{comentario2}, the desorption term in the radiative equation has been measured to be much smaller than the adsorption one and can be neglected. The adsorption term is the product of the ion flux close to the surface and the probability $p_I$ that an ion will be neutralised on the surface. The ion flux itself is proportional to the flux of excited atoms and to their two-photon ionization rate:
\begin{equation}
\frac{dN_I}{dt}= p_I \; \sigma_2\; \sigma_1(\delta)\; g(\delta) \frac{I^3}{(\hbar \omega)^3} \sqrt{\frac{k_B T_w}{2 \pi m}}  \tau^2 \; n_A ,
\label{adionflux}
\end{equation}
which is consistent with our observation that this contribution is proportional to the atomic density and has a cubic dependence on the laser intensity for constant $p_I$. In eq.~(\ref{adionflux}), $\sigma_2$ is a generalised cross section for two-photon absorption, $\tau$ is the lifetime of excited atoms, $\delta$ is the laser detuning in relation to the D$_2$ central frequency and $\sigma_1(\delta)$ is the cross section for absorption by a ground-state atom. In $\sigma_1(\delta)$ the Doppler-broadening effect of atoms in motion is taken into account. The factor $g(\delta)$ incorporates the resonant dip due to the surface quenching of excited atoms \cite{Zajonc1981} (see fig.~\ref{fig:intensity}(b)).  

	A simple model of the interaction between a solid surface and a singly charged ion makes explicit the dependence of the probability $p_I$ \cite{VanWunnik1983} on the coverage-dependent work function of the substrate \cite{Brison2007} $\phi(\theta) = \phi_0 - \gamma \theta$ and the consequent threshold behaviour.

\begin{equation}
p_I= \frac{\displaystyle \sum_{\vec{k}}\left|T_{\vec{k}a}\right|^2 L(\epsilon_{\vec{k}}, \epsilon_a, \Delta) f(\epsilon_{\vec{k}}, \epsilon_F, T_w)}{\displaystyle \sum_{\vec{k}}\left|T_{\vec{k}a}\right|^2 L(\epsilon_{\vec{k}}, \epsilon_a, \Delta)}
\label{equ:psum}
\end{equation} 
	
$p_I$ is the product of $T_{\vec{k}a}= \left\langle a |V|\vec{k}\right\rangle$, the transition probability of an electron from a surface level $|\vec{k}\rangle$ to a vacant atomic level $|a\rangle$, the surface Fermi distribution $f(\epsilon_{\vec{k}}, \epsilon_F, T_w)$ at temperature $T_w$ and the atomic lineshape $L(\epsilon_{\vec{k}}, \epsilon_a, \Delta)$, of width $\Delta$, that we approximate by a Dirac delta function at the energy $\epsilon_a$ of the vacant atomic level. With such simplifications we obtain:
	
\begin{equation}
p_I = \frac{1}{1+\textrm{exp}\left[\left(\phi_0 -|\epsilon_a|\right)-\gamma \theta\right]/k_B T_w},
\label{equ:p}
\end{equation}
where $\phi_0$ is the work function of the clean surface and $\gamma$ is a proportionality coefficient. The range of Cs coverage in which $p_I$ significantly varies is very narrow (see fig.~\ref{fig:estates}(c)). The order of magnitude of the difference ($\phi_0-|\epsilon_a|$) is a few eV for an alkali atom close to a clean metallic surface \cite{Borisov1999}. For our experiment, $p_I$ can be assumed to be a constant (saturated value) almost immediately above the threshold density, so that the film growth rate varies linearly with the atomic density as observed (figs.~\ref{fig:density}(a) and \ref{fig:density}(b)). We thus interpret all of the features observed during the Cs film growth on our cell's window. Fits of the growth rate as a function of the atomic density, with eqs.~(\ref{equ:rate})-(\ref{equ:p}), are shown in figs.~\ref{fig:density}(a) and \ref{fig:density}(b). The fixed parameters in each case are the experimental values of the intensity, of the window temperature and of the atomic density, as well as $\gamma$ = 20 ev (from \cite{Brison2007}) and ($\phi_0-|\epsilon_a|$)= 4 eV (from \cite{Brause1997}) in eq. (\ref{equ:p}). The adjustable parameter is $\sigma = (dN_I/dt)/(p_I n_A I^3)$. Using a typical measured rate of film growth, $dL/dt \approx$ 4 $\times 10^ {-3}$ nm.s$^ {-1}$, we estimate the adsorption rate of ions at these intensities to be on the order of 10$^{11}$ atoms/s on the illuminated surface. The fits yield values of $\sigma$ on the order of 10$^{-2}$ J$^{-3}$.s$^2$.cm$^7$. The functional dependence of the fits on the parameters is consistent with the experimental measurements. A detailed quantitative analysis of the process, and particularly the estimation of the effective ion yield in the vapour, depends on a systematic study of the function $g(\delta)$ for a given material of the window.

In conclusion, we have identified the sequence of vapour and surface processes leading to the formation and growth of a light-induced thin metallic film at the interface between a transparent dielectric solid and an alkaline vapour. Preliminary results show that the structure is very stable, provided that the window temperature is quickly lowered after the pump laser is switched off. In addition to its technological appeal \cite{Meschede2003}, this film growth technique also provides a method of characterizing various aspects of the interaction between a dielectric surface and gas phase atoms, particularly the adsorption time and energy and the caesiation-induced lowering of the work function in dielectric substrates. Further spatial study of the film growth may also provide insights into effects neglected in this work, such as dark- or light-induced diffusion on the surface.\\

\acknowledgments
We acknowledge support from FINEP, CNPq and CAPES (Brazil). We warmly acknowledge J.W.R. Tabosa for carefully reading the manuscript.


\begin{thebibliography}{0}


\bibitem{Obrecht2007A} 
  Obrecht J. M.,  Wild R. J. and Cornell E. A., Phys. Rev. A \textbf{75}, 062903 (2007).
  
\bibitem{Happer1972}
  Happer W., Rev. Mod. Phys. \textbf{44}, 169 (1972).
  
\bibitem{Bouchiat1999}
  Bouchiat M.A., Gu\'{e}na J., Jacquier Ph., Lintz M. and Papoyan A.V., Applied Physics B: Lasers and Optics \textbf{68}, 1109 (1999).

\bibitem{Hatakeyama2006}
  Hatakeyama A., Wilde M. and Fukutani K., e-Journal of Surface Science and Nanotechnology \textbf{4}, 63(2006).
    
\bibitem{Lima2000}
  Lima E.G., Chevrollier M., Di Lorenzo O., Segundo P.C. and Ori\'{a} M., Phys. Rev. A \textbf{62}, 013410 (2000).
  
\bibitem{deSilans2006}
  Passerat de Silans T., Farias B., Ori\'{a} M. and Chevrollier M., Appl. Phys. B \textbf{82}, 367 (2006).

\bibitem{leKien2007}
  Fam Le Kien, Dutta Gupta S. and Hakuta K., Phys. Rev. A \textbf{75}, 032508 (2007).
  
\bibitem{Gunther2005}
  G\"unther A., Kraft S., Kemmler M., Koelle D., Kleiner R., Zimmermann C. and Fort\'agh J., Phys. Rev. Lett. \textbf{95}, 170405 (2005).
  
\bibitem{Jha2011}
  Jha P.K., Dorfman K.E., Yi Z., Yuan L., Sautenkov V.A., Rostovtsev  Y.V., Welch  G.R., Zheltikov A.M. and Scully M.O., Appl. Phys. Lett. \textbf{101}, 091107 (2012).
  
\bibitem{Afanasev2008}
  Afanas'ev A.E., Melent'ev P.N. and Balykin V.I., Bulletin of the Russian Academy of Sciences: Physics \textbf{72}, 664 (2008).
  
\bibitem{Afanasiev2007}
  Afanasiev A., Melentiev P. and Balykin V., JETP Letters \textbf{86}, 172 (2007).
  
\bibitem{comentario1}
Here we study the film starting process. The regime of thick film will be discussed elsewhere.

\bibitem{Tam1975}
Tam A., Moe G. and Happer W., Phys. Rev. Lett. \textbf{35}, 1630 (1975).
    
\bibitem{Palik1985}
  Palik E. D., Handbook of Optical Constants of Solids (Academic Press, 1985).
  
\bibitem{Zajonc1981}
  Zajonc A.G. and Phelps A.V., Phys. Rev. A \textbf{23}, 2479 (1981).
   
\bibitem{deBoer1953}
  de Boer J.H., The dynamical character of adsorption (Oxford University Press, 1953).
   
\bibitem{Hoinkes1980}
  Hoinkes H., Review of Modern Physics \textbf{52}, 933 (1980).
  
\bibitem{Norskov1990}
  Norskov J. K., Rep. Prog. Phys. \textbf{53}, 1253 (1990).
  
\bibitem{comentario2}
 We estimated the thermal desorption rate at a given temperature of the window by measuring the variation of the equilibrium caesium adatoms density (Eq. (\ref{equ:Nab})) as a function of the vapour density, all other parameters kept constant, and assuming a sticking coefficient $p_A$ equal to unity. The light-induced film desorption rates are obtained by measuring the decreasing of the film thickness when the laser is turned off, all other conditions being the same.
  
\bibitem{Laidler1996} 
  Laidler K.J., Pure \& Appl. Chem. \textbf{68}, 149 (1996).

\bibitem{Yu1984}
  Yu M.L., Phys. Rev. B \textbf{29}, 2311 (1984).
  
\bibitem{Brause1997}
  Brause M., Ochs D., G\"{u}nster J., Mayer Th., Braun B., Puchin V., Maus-Friedrichs W. and Kempter V., Surface Science \textbf{383}, 216 (1997).
  
\bibitem{Kamins1968a}
  Kamins T.I., J. Appl. Phys. \textbf{39}, 4529 (1968).
  
\bibitem{Gurney1935}
  Gurney R. W., Phys. Rev. \textbf{47}, 479 (1935).

\bibitem{Brison2007}
  Brison J., Mine N., Poisseroux S., Douhard B., Vitchev R.G. and Houssiau L., Surface Science \textbf{601}, 1467 (2007).
    
\bibitem{Winter2002}
  Winter H., Physics Reports \textbf{367}, 387 (2002).
    
\bibitem{VanWunnik1983}
  Van Wunnik J.N.M., Brako R., Makoshi K. and Newns D.M., Surface Science \textbf{126}, 618 (1983).
  
\bibitem{Borisov1999}
  Borisov A.G., Kazansky A.K. and Gauyacq J.P., Surface Science \textbf{430}, 165 (1999).
  
\bibitem{Meschede2003}
  Meschede D. \and Metcalf H., J. Phys. D: Appl. Phys. \textbf{36}, R17 (2003).
  
\end{thebibliography}
\end{document}